\pdfoutput=1
\documentclass[%
 aip,
 amsmath,amssymb,
 reprint,%
]{revtex4-1}

\usepackage{graphicx}%
\usepackage{dcolumn}%
\usepackage{bm}%
\usepackage{xcolor}
\usepackage[utf8]{inputenc}
\usepackage[T1]{fontenc}
\usepackage{mathptmx}
\usepackage{etoolbox}
\usepackage[hidelinks,colorlinks=true,linkcolor=blue,citecolor=blue]{hyperref}

\makeatletter
\def\@email#1#2{%
 \endgroup
 \patchcmd{\titleblock@produce}
  {\frontmatter@RRAPformat}
  {\frontmatter@RRAPformat{\produce@RRAP{*#1\href{mailto:#2}{#2}}}\frontmatter@RRAPformat}
  {}{}
}%
\makeatother

\begin{document}

\preprint{AIP/123-QED}

\title{Successful irradiation campaign on PRobe far-Infrared Mission for Astrophysics (PRIMA)/PRIMAger Kinetic Inductance Detectors with the Detector irRAdiation Cryogenic faciLity for Astrophysics (DRACuLA)}

\author{Valentin Sauvage}
\email{valentin.sauvage@universite-paris-saclay.fr}
\affiliation{Institut d'Astrophysique Spatiale, Université Paris-Saclay, CNRS, Orsay, France}

\author{Anaïs Besnard}
\affiliation{Institut d'Astrophysique Spatiale, Université Paris-Saclay, CNRS, Orsay, France}

\author{Giulia Conenna}%
\affiliation{SRON, Space Research Organisation Netherlands, Niels Bohrweg 4, 2333 CA Leiden, and Landleven 12, 9747 AD Groningen, The Netherlands}

\author{Kenichi Karatsu}
\affiliation{SRON, Space Research Organisation Netherlands, Niels Bohrweg 4, 2333 CA Leiden, and Landleven 12, 9747 AD Groningen, The Netherlands}

\author{Stephen J.C. Yates}
\affiliation{SRON, Space Research Organisation Netherlands, Niels Bohrweg 4, 2333 CA Leiden, and Landleven 12, 9747 AD Groningen, The Netherlands}

\author{Lorenza Ferrari}
\affiliation{SRON, Space Research Organisation Netherlands, Niels Bohrweg 4, 2333 CA Leiden, and Landleven 12, 9747 AD Groningen, The Netherlands}

\author{Bruno Maffei}
\affiliation{Institut d'Astrophysique Spatiale, Université Paris-Saclay, CNRS, Orsay, France}

\begin{abstract}
DRACuLA (\textbf{D}etector ir\textbf{RA}diation \textbf{C}ryogenic faci\textbf{L}ity for \textbf{A}strophysics) is a mobile dilution refrigerator platform developed at the Institut d'Astrophysique Spatiale (IAS) to expose sub-Kelvin detectors to particle beams at their nominal operating temperature, in the range 50--300~mK. We report on its design, beam-line integration at the Particle Therapy Research Center (PARTREC) in Groningen, and the operational performance achieved during the September~2025 irradiation campaign on Kinetic Inductance Detector (KID) arrays developed by SRON for the PRIMA mission. The detector samples were maintained at 120~mK throughout a 12-hour proton irradiation run at 184~MeV. The scientific results of this campaign are reported in a companion paper.
\end{abstract}

\maketitle

\section{Introduction}

The next generation of space observatories, such as PRIMA~\cite{Glenn2025}, Athena~\cite{Nandra2013}, and LiteBIRD~\cite{Hazumi2020}, relies on increasingly sensitive superconducting detectors, including Kinetic Inductance Detectors (KIDs) and Transition Edge Sensors (TES), operating at sub-Kelvin temperatures. In the space environment, cosmic-ray particles interact with detector materials through ionisation and atomic displacement, giving rise to two classes of effect: transient signal glitches that degrade data quality, as extensively documented during the Planck HFI mission~\cite{Ade2011}, and potential permanent degradation of detector performance through accumulated radiation damage~\cite{DalBo2026, Besnard2026b}.

Qualifying KIDs against these effects prior to launch is therefore mandatory. Early irradiation campaigns, such as that reported by Karatsu et al.~\cite{Karatsu2016} using a 160~MeV proton beam at room temperature, demonstrated no significant damage at doses of $\sim$10~krad. However, radiation-induced lattice defects are known to be mobile at room temperature, leading to annealing of the damage prior to cooling the devices to their operating conditions for characterization. In-situ irradiation at cryogenic temperatures is therefore essential to capture the true impact of particle hits on detector performance.

To address this need, IAS has developed DRACuLA (\textbf{D}etector ir\textbf{RA}diation \textbf{C}ryogenic faci\textbf{L}ity for \textbf{A}strophysics), a mobile dilution-refrigerator platform designed to expose sub-Kelvin detectors to particle beams at their nominal operating temperature. The cryostat was commissioned in 2018~\cite{Janssen2018Commissioning} and subsequently equipped with the readout electronics, wiring, filters and thermometry required for detector testing. Its design was driven by the following requirements: a base temperature as low as 10~mK, a cooling power of 500~$\mu$W at 100~mK, compatibility with any accelerator beam line, and a sample space large enough to accommodate detector arrays or part of a focal plane. IAS has performed three irradiation campaigns with DRACuLA: a first in September~2022 at ALTO (IJCLab) in Orsay on composite bolometers, a second in May~2024~\cite{Besnard2024} at the same facility on TES detectors developed by NIST for LiteBIRD, and most recently in September~2025 at the Particle Therapy Research Center (PARTREC) in Groningen, the Netherlands, on KID arrays developed by SRON for PRIMA, the subject of this article.

This article describes the design, beam-line integration, and operational performance of DRACuLA during the September~2025 campaign, the first deployment of the facility away from its home laboratory and on a high-energy proton beam line. The scientific results obtained, namely the performance of KID arrays developed for the PRIMAger focal plane of the PRIMA mission following irradiation to the ionising dose specified for the mission, are reported in the companion paper by Besnard et al.~\cite{Besnard2026}.

\section{Experiment Overview}

\subsection{The DRACuLA Cryogenic Facility}

DRACuLA was conceived around a single core requirement: to expose any sub-Kelvin detector to a particle beam while it operates at its nominal temperature. Modularity is the primary design driver; the facility must interface with any particle accelerator without requiring a fundamental redesign. As in its original design~\cite{Janssen2018Commissioning}, the platform is mounted on a compact frame equipped with wheels, making it transportable between accelerator facilities, and the vacuum can is suspended from the frame through bellows to limit the transmission of vibrations. It provides 24 RC/RF low-pass filtered readout lines (QDevil QFilter-II) and four superconducting coaxial wires spanning from room temperature down to the cold plate, together with four independent temperature measurement bridges.

The cryogenic core of DRACuLA is a Bluefors LD400 dilution refrigerator. The thermal architecture comprises four successive cold stages: 50~K and 4~K (provided by a Cryomech PT415-RM pulse tube cooler), followed by the 1~K and 100~mK plates, each enclosed by the corresponding radiation shield (with the exception of the 100~mK stage).

\begin{figure}
    \centering
    \includegraphics[width=\linewidth]{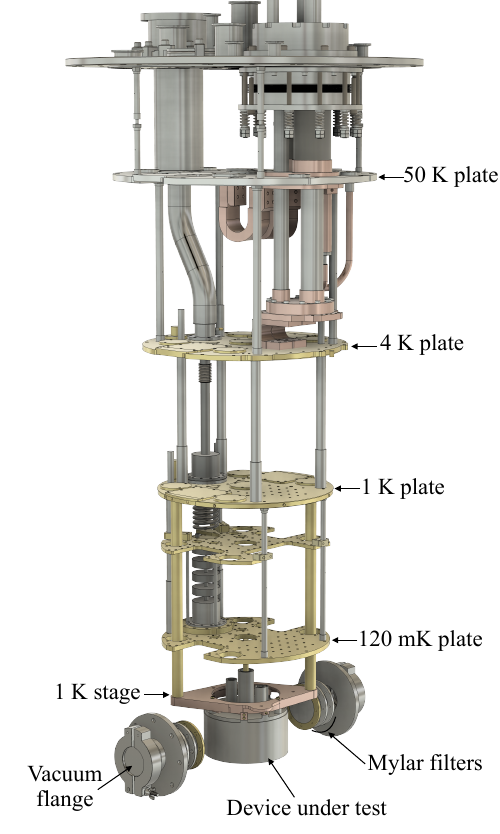}
    \caption{CAD view of the experimental setup (Device under test within the DRACuLA cryostat).}
    \label{fig:DRACuLA_anatomy}
\end{figure}

To allow particle beam access to the cold sample, four KF-50 ports are machined through the cryostat vacuum can and all inner shields at angles of 0$^\circ$, 45$^\circ$, 90$^\circ$, and 180$^\circ$. This four-port geometry provides flexibility in beam-line coupling without any modification to the cryostat itself (see Fig.~\ref{fig:DRACuLA_anatomy}). 

\subsection{Beam Line Integration at PARTREC}

The September~2025 campaign was carried out at the Particle Therapy Research Center (PARTREC), located at the University Medical Center Groningen (UMCG), the Netherlands~\cite{Gerbershagen2022}. PARTREC was selected for its proton beam energy range, location and availability. The facility delivers proton beams at energies up to 190~MeV, well-suited to simulate the cosmic-ray proton spectrum expected at the L2 Lagrange point over the mission lifetime.

The proton beam energy at the accelerator exit was 184~MeV. As the beam traverses the successive layers of the experimental setup: the collimator, the aluminum vacuum flange, the aluminized mylar filters on each thermal shield, the Niobium/CryoPerm shielding of the light-tight box and the gold-plated copper shield, it undergoes progressive energy loss. The residual energy at each layer was estimated by GEANT4 simulation~\cite{Agostinelli2003}, yielding a proton energy from 184~MeV to approximately 150~MeV at the detector level.

\section{Campaign Execution}

The DRACuLA facility was transported from IAS in Orsay to PARTREC in Groningen and installed in a dedicated side room for pre-campaign integration and commissioning tests. The PID control parameters were optimized during this phase using the newly assembled setup. Two types of KID arrays, absorber coupled and antenna coupled, were characterized prior to irradiation to establish a reference dataset adapted to the environmental conditions of the beam hall, which differ from the SRON laboratory setup. The full experiment was subsequently moved into the beam hall as soon as the facility granted access, aligned with the proton beam line, and cooled to a base temperature of 120~mK. 

It should be noted that this campaign also enabled the characterization, under identical conditions, of two-stage SQUIDs operating at 1~K for LiteBIRD, as well as LC resonators at 120~mK for Athena X-IFU. These devices were specifically integrated and positioned within the cryostat so as not to interfere with the primary objective of the campaign. 

\subsubsection{Alignment}

\begin{figure}
    \centering
    \includegraphics[width=\linewidth]{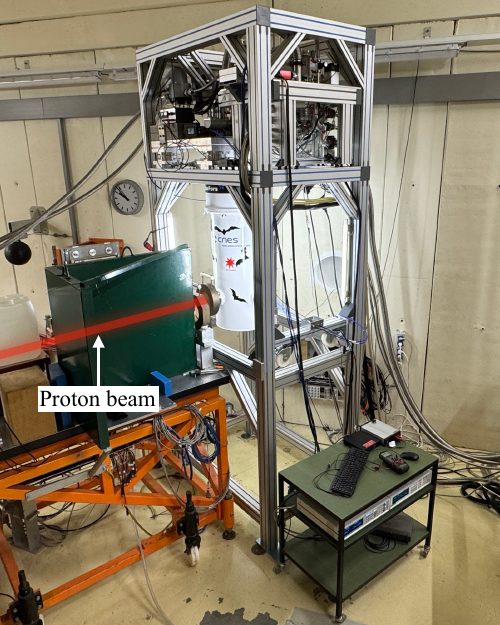}
    \caption{The cryostat mounted on its support frame in front of the beam line at PARTREC, raised to 1.5~m to align the beam axis with the cryostat windows.}
    \label{fig:overall_picture}
\end{figure}

\begin{figure}
    \centering
    \includegraphics[width=\linewidth]{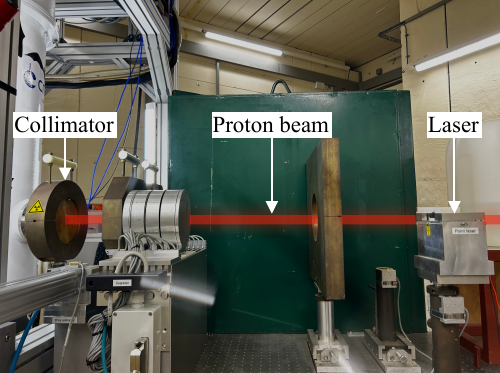}
    \caption{DRACuLA in front of the proton beam line with the collimator
    reducing the beam cross-section to 20$\times$20~mm. The alignment laser
    is visible along the beam axis and was used to position the cryostat
    windows with respect to the beam.}
    \label{fig:alignment}
\end{figure}

Mechanical alignment of DRACuLA with the beam line was achieved using a supplementary support frame that raises the cryostat to 1.5~m height, where the beam axis is located, as shown in Fig.~\ref{fig:overall_picture}. A laser placed along the nominal beam axis served as an alignment reference to position the cryostat windows accurately with respect to the beam (Fig.~\ref{fig:alignment}). A collimator was installed outside the cryostat, in front of the KF-50 beam port, to reduce the beam cross section to 20$\times$20~mm at the detector plane, matching the dimensions of the KID demonstrator arrays while limiting the beam heat load on the cold stages.

\subsubsection{Experimental Setup}

The device under test (DUT) is enclosed in a light-tight box to prevent stray photons from reaching the detectors during dark measurements. Magnetic shielding is provided by a combination of Niobium and CryoPerm enclosures surrounding the sample holder. During the 2025 campaign, both the absorber-coupled and antenna-coupled KID arrays were mounted back-to-back in the sample holder and thermalized to the 120~mK stage~\cite{Besnard2026}. The readout chain comprised superconducting coaxial feedthroughs, a cold low-noise amplifier (LNA) at the 4~K stage, a room-temperature LNA outside the cryostat, and a single-tone homodyne readout based on a NI PXI-1036 chassis with a NI PXI-5922 digitizer sampling at 50~kHz~\cite{Besnard2026}.

\subsubsection{Thermal Performance During Irradiation}

\begin{figure}
    \centering
    \includegraphics[width=\linewidth]{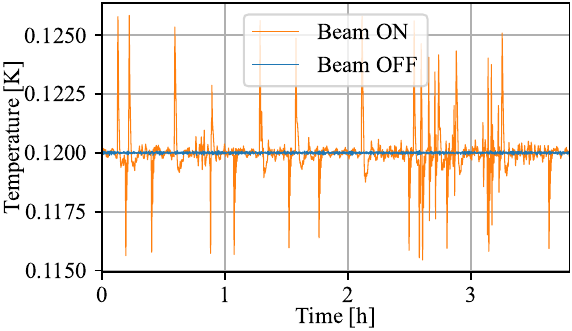}
    \caption{Temperature evolution of the 120~mK stage over a 4-hour window before beam operation (Beam OFF) and during irradiation (Beam ON). Temperature overshoots above 120~mK are induced by fluctuations in beam flux.}
    \label{fig:temperature_120mK}
\end{figure}

\begin{figure}
    \centering
    \includegraphics[width=\linewidth]{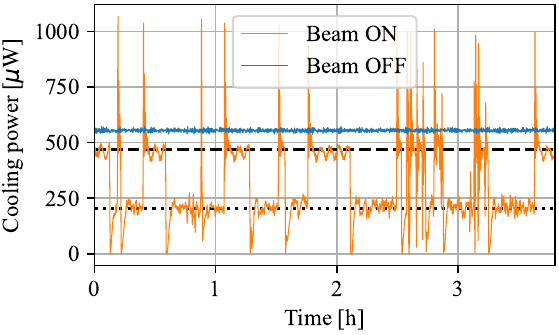}
    \caption{Cooling power at the 120~mK stage over a 4-hour window before (Beam OFF) and during irradiation (Beam ON). The three regimes correspond to nominal operation (blue), single turbo-molecular pump operation (\textbf{-~-}), and active beam time (\textbf{..}). Fluctuations during Beam ON are driven by beam flux variations.}
    \label{fig:cooling_power_120mK}
\end{figure}

Temperature stability at the detector level was monitored by a Lakeshore~372 AC resistance bridge reading a RuO$_2$ thermometer calibrated down to 30~mK, installed at the 120~mK thermal link. A 120~$\Omega$ resistive heater mounted directly on the cold plate provided the heating power for the PID controller (see note~\cite{note1}). Prior to beam operation, a thermal stability of 120~mK~$\pm$~33~$\mu$K was achieved (Fig.~\ref{fig:temperature_120mK}, Beam OFF). Under irradiation, the stability deteriorated to 120~mK~$\pm$~695~$\mu$K over the full beam period (Fig.~\ref{fig:temperature_120mK}, Beam ON). This degradation is primarily attributable to temperature overshoots induced by fluctuations and interruptions in beam intensity, which drove the PID controller, optimized without beam load, outside its optimized range, as visible in Fig.~\ref{fig:cooling_power_120mK}. Since beam time is costly, the proton flux was deliberately increased to reach the target dose quickly while staying within the capability of the cryostat, leaving only a small cooling margin. The beam intensity was not recorded and could not be retrieved from the facility.

Three distinct cooling power regimes are identifiable in Fig.~\ref{fig:cooling_power_120mK}. The first corresponds to nominal Beam OFF operation, with 554~$\mu$W of available cooling power. The second, at 467~$\mu$W, follows the failure of one of the two turbo-molecular pump controllers, which reduced the cooling power at the detector level. The third regime, at 203~$\mu$W, covers the active beam period and reflects the combination of the reduced pumping capacity and the additional heat load deposited by the beam. The cooling power is here the power dissipated by the PID heater, measured in a four-wire configuration. A Bethe--Bloch estimate of the energy deposited by the protons in the materials attached to the 100~mK and 1~K stages gives a direct beam heat load of about 40~$\mu$W and 45~$\mu$W, respectively. This estimate is a lower bound, as it neglects scattered protons, beam halo and secondary particles, while the measured reduction of 264~$\mu$W is an upper bound, since the time-averaged heater power underestimates the available cooling power when the PID output drops to zero during temperature overshoots.

\begin{figure}
    \centering
    \includegraphics[width=\linewidth]{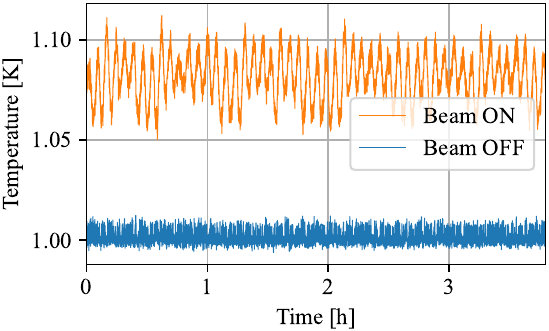}
    \caption{Temperature of the 1~K stage before (Beam OFF) and during (Beam ON) irradiation. The available cooling power was sufficient to maintain 1~K before beam operation but not during irradiation due to the combined effects of beam heating and reduced pumping capacity.}
    \label{fig:temperature_1K}
\end{figure}

\begin{figure}
    \centering
    \includegraphics[width=\linewidth]{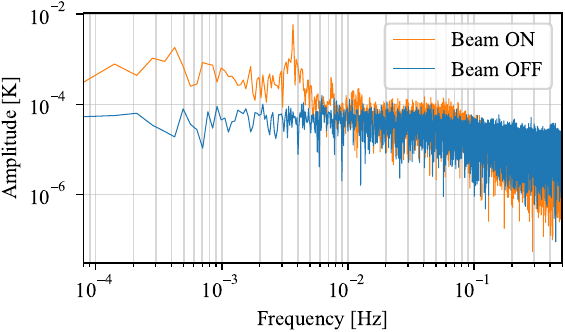}
    \caption{Discrete Fourier Transform of the 1~K stage temperature before and during irradiation. A peak near 34~mHz, corresponding to the natural oscillation frequency of the dilution system, appears during Beam ON and is absent during Beam OFF, where it is suppressed by the PID controller.}
    \label{fig:dft_1K}
\end{figure}

Temperature at the 1~K shield stage was monitored by a RuO$_2$ thermometer calibrated to 300~mK, read by a Lakeshore~370 resistance bridge, with a 3~k$\Omega$ heater providing PID-controlled stabilisation at 1~K. Before irradiation, the stage was maintained at 1~K~$\pm$~2.7~mK. During irradiation, the available cooling power was insufficient to hold the stage at 1~K; the temperature rose to approximately 1.081~K with a stability of $\pm$~11.7~mK (Fig.~\ref{fig:temperature_1K}). This instability resulted from two concurrent contributions: the direct heat load of the beam on the 1~K shield, and the reduced pumping capacity on the still caused by the turbo-molecular pump failure. Under these conditions, the natural oscillation of the dilution system at approximately 34~mHz (normally suppressed by the PID) became visible in the temperature spectrum (Fig.~\ref{fig:dft_1K}).

The irradiation run lasted approximately 12 hours at a proton flux of $\sim$2$\times$10$^6$~protons/cm$^2$/s at detector level. The corresponding fluence, $\sim$8.4$\times$10$^{10}$~protons/cm$^2$, yields a total dose of 5.7~krad in the aluminium film of the detectors (ionising dose of 3.5$\times$10$^{11}$~MeV/g, displacement damage dose of 1.8$\times$10$^{8}$~MeV/g)~\cite{Besnard2026}. The ionising dose matches the 5.76~krad specified for the 64-month PRIMA mission~\cite{PRIMAERD}, while the displacement damage dose is about six times the mission value computed by Kane et al.~\cite{Kane2026}. The beam was periodically turned off, and the detector performance measured as in~\cite{Besnard2026}, before restarting the beam to complete the radiation dose. 

Immediately after the beam was definitively switched off, a considerable increase of the glitch rate was observed in the detector time-ordered data: approximately 15\% of data points were flagged as glitches in the post-irradiation dataset, compared to $\sim$1\% before irradiation. The primary source was identified as radio-activation of the gold-plated copper sample enclosure, with $^{64}$Cu ($\beta^+$ and $\beta^-$ decay, $t_{1/2}=12.7$~h) as the dominant activated isotope. Post-irradiation science measurements were therefore delayed by 24 hours to allow the $^{64}$Cu activity to decay to a level compatible with reliable data acquisition. The detector measurements, reported in the companion paper~\cite{Besnard2026}, show no significant change in quasiparticle lifetime and dark responsivity after irradiation, while the observed increase of the dark NEP is tentatively attributed to incomplete deglitching of the post-irradiation data.

\section{Improvements for the Next Campaign}
\label{sec:improvements}

The 2025 campaign highlighted several areas for improvement in future operations of DRACuLA. During the 12-hour beam run, the facility encountered two significant incidents.

The controller of one of the two turbo-molecular pumps failed while the proton flux was being raised to its nominal value, with no replacement available on-site. Since the two pumps operate in parallel, DRACuLA continued operating with a single pump at the cost of reduced cooling power at the mixing chamber and the still (1~K plate). The associated power surge triggered the fuses of the experimental room, causing a beam interruption and requiring an in situ intervention. The primary cryostat operation was not compromised, and DRACuLA was restored to near-full functionality within 15 minutes; the cold stage temperature remained at 120~mK.

The second incident was the failure of the DRACuLA control computer approximately one hour before the scheduled end of the beam run, again requiring an in-situ intervention. The manufacturer later confirmed that the electronic board of the pump controller, housed in the cryostat frame less than 80~cm from the beam axis, was damaged, while the computer restarted normally. In the absence of dosimetry at the location of the electronics, both events are most plausibly due to stray radiation from the beam. Such effects were not observed at ALTO, where the beam energy ($\leq$22~MeV) is too low for the beam to cross the cryostat.

For future campaigns, spare units of the turbo-molecular pump controller, of the control computer and of the temperature measurement bridges will be available on site. This mitigation concerns only the room-temperature electronics and requires no modification of the cryostat.

\section{Conclusion}

We have presented DRACuLA, the mobile cryogenic irradiation facility developed at IAS for in-situ characterization of sub-Kelvin detectors under particle beam exposure. Its modular design, built around a Bluefors LD400 dilution refrigerator with four-angle KF-50 beam access ports, provides 500~$\mu$W of cooling power at 100~mK and has been successfully operated at two different accelerator facilities: ALTO (IJCLab) in Orsay and PARTREC in Groningen.

Over three campaigns since 2022, covering composite bolometers, TES detectors, and KID arrays, DRACuLA has demonstrated robust thermal performance under beam exposure at temperatures as low as 100~mK, and has yielded the first cryogenic in-situ total-dose irradiation result for KIDs at the PRIMA mission dose level. The KID arrays developed by SRON for the PRIMAger focal plane of the PRIMA mission showed no significant change in quasiparticle lifetime and dark responsivity at 5.7~krad~\cite{Besnard2026}, supporting their qualification for future space missions. 

The DRACuLA facility is now available to the community for Technology Readiness Level (TRL) qualification of sub-Kelvin detector technologies intended for future space and balloon-borne observatories.

\section*{Author declarations}
\subsection*{Conflict of interest}
The authors have no conflicts to disclose.
\subsection*{Author Contributions}
\noindent
\textbf{Valentin Sauvage:} writing - original draft preparation (lead); writing - review and editing (equal); resources - instrumentation (equal); investigation - managing cryogenic facility (lead). 
\textbf{Anaïs Besnard:} writing - review and editing (equal); resources - instrumentation (equal); investigation - managing cryogenic facility (supporting). \textbf{Giulia Conenna:} writing - review and editing (lead); resources - detectors (equal); \textbf{Kenichi Karatsu:} writing - review and editing (equal); resources - detectors (equal); \textbf{Stephen J.C. Yates:} writing - review and editing (equal); resources - detectors (equal); \textbf{Lorenza Ferrari:} writing - review and editing (equal); conceptualization (lead); resources - detectors (equal); project administration (lead);
\textbf{Bruno Maffei:} writing - review and editing (equal); funding acquisition (lead); resources - instrumentation (lead). 

\section*{Data availability}

Data available on request from the authors.

\begin{acknowledgments}
    This facility has been co-funded by the French National Space Agency (CNES) and the DIM-ACAV research programme of the Ile-de-France region, with participation of the OSUPS (Observatoire des Sciences de l'Univers Paris-Saclay). The authors thank the staff of PARTREC and SRON at the University of Groningen for their support during the irradiation campaign.
\end{acknowledgments}

\bibliographystyle{aipnum4-1}
\bibliography{bibliography}

\begin{thebibliography}{15}
\bibitem [{not()}]{note1}%
  \BibitemOpen
  \href@noop {} {}\bibinfo {note} {Due to a lack of space, the heater was not
  placed closed to the thermometer, resulting in a degraded thermal stability
  due to the distance inducing a time delay between the heating and the
  thermometer response. The achieved thermal stability was sufficient for this
  experiment}\BibitemShut {NoStop}%
\end{thebibliography}

\end{document}